\documentclass[onecolumn, a4size, 11pt]{article}
\usepackage{amsfonts}
\usepackage{spconf,color}
\usepackage{amstext,amssymb,epsfig,psfrag,graphicx,cite,footnote,amsmath}
\usepackage{epstopdf}
\usepackage{mathdots}
\usepackage{arydshln}
\usepackage{mathrsfs}

\title{Joint Cooperative Computation and Interactive Communication for  Relay-Assisted Mobile Edge Computing \vspace{-0.5em} }
\DeclareFixedFont{\myfont}{OT1}{cmr}{m}{n}{8pt}
\name{Xihan Chen{\small $~^{\ast}$}, Qingjiang Shi{\small $~^{\dagger}$}, Yunlong Cai{\small $~^{\ast}$}, Minjian Zhao{\small $~^{\ast}$},  and Mingjie Zhao{\small $~^{\ddagger}$}  \vspace{-0.5em}
\address{\myfont{$^{\ast}$\,Department of ISEE, Zhejiang University,  Hangzhou, China} \\
\myfont{$^{\dagger}$\,College of EIE, Nanjing University of Aeronautics and Astronautics, Nanjing
, China}\\
\myfont{ $^{\ddagger}$\,School of Info. and Sci. Tech., Zhejiang Sci-Tech University, Hangzhou, China}\\   \myfont{\,chenxihan@zju.edu.cn, \,qing.j.shi@gmail.com,}  \,ylcai@zju.edu.cn,  \, mjzhao@zju.edu.cn, \,zhaomjx@163.com \vspace{-0.5em}}}
\begin{document} \linespread{0.75} \ninept

\maketitle
\vspace{-0.5em}
\begin{abstract}
To realize cooperative computation and communication in a relay mobile edge computing system, we develop a hybrid relay forward protocol, where we seek to balance the execution delay and network energy consumption. The problem is formulated as a nondifferentialbe  optimization problem which is nonconvex with  highly coupled constraints. By exploiting the problem structure, we propose a lightweight algorithm based on inexact block coordinate descent method. Our results show that the proposed algorithm exhibits much faster convergence as compared with the popular concave-convex procedure based
algorithm,  while achieving good performance.
\end{abstract}
 \vspace{-.5em}
\section{Introduction}
 \vspace{-.5em}
\label{sec:intro}
The mobile edge computing (MEC) has been considered as a new network architecture that enables cloud computing capabilities and IT service environment at the edge of network. This  architecture has the potential to significantly reduce latency, avoid congestion and prolong the battery lifetime of mobile devices, by pushing data intensive tasks towards the edge and locally processing data in proximate MEC server\cite{1}.

Recently, MEC has gained a lot of interest \cite{2,3,4,5,6}. In \cite{3} and \cite{4}, the authors derived the optimal resource allocation solution for a single-user MECO system with multiple elastic tasks to minimize the average execution latency of all tasks under power constraints. You \emph{et al.} \cite{2} investigated the optimal resource and offloading decision policy to minimize the weighted sum mobile energy consumption under the computation latency constraint in a multiuser  TDMA/OFDMA MEC system. The work \cite{5} considered the joint optimization of radio and computational resources for computation offloading in a dense
deployment scenario, in the presence of intercell interference. Wang \emph{et al.} \cite{6} presented an ADMM-based decentralized algorithm for computation offloading, resource allocation and internet content caching optimization in heterogeneous wireless cellular networks with MEC.

The aforementioned works focused on scenario where mobile terminals offload their computational tasks to MEC server and then the latter feedbacks the results to mobile terminals.
In this paper, differently from the current MEC works, we consider a cooperative computing and interactive communication system as shown in Fig. 1 where user A wants to share his computational result with  user B with the aid of a \textbf{R}elay equipped with an \textbf{ME}C \textbf{S}erver (abbreviated as R-MES). 
To reduce the execution delay and the energy consumption, we seek to minimize a weighted sum of the execution delay and network energy consumption subject to radio and computation resource constraints at users and R-MES. The problem is a nonconvex nondifferentiable problem with highly coupled constraints. By exploiting the problem structure as fully as possible, we adopt smooth approximation and inexact block coordinate descent method to address the difficulties arising from the nondifferentiability, nonconvexity and constraint coupling.
 Simulation results show that the proposed algorithm significantly improves the speed of convergence while achieve good performance as compared with the concave-convex procedure (CCCP)-based algorithm.


 \vspace{-.5em}
\section{System model and problem statement}

\begin{figure}[t]
\centering
\includegraphics[width=3.00in,height=1.0in]{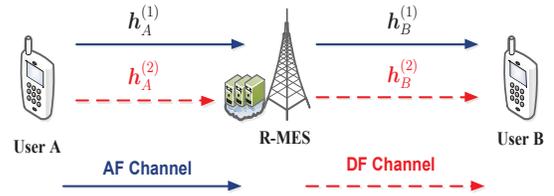}
\caption{User A shares computation results with User B assisted by an MEC relay.}
\label{fig:side:a}
 \vspace{-2em}
\end{figure}

As shown in Fig. 1, we consider a computational result sharing (CRS) system where user A wants to share its computational result with user B over a relay channel. It is assumed that the relay is equipped with an MEC server and it has the capability to help user A process the tasks. Given the computational tasks, user A decides whether to locally process the tasks or offload all its tasks or part of its tasks to the R-MES. To support CRS,  we propose a hybrid relay forward (HRF) protocol, where, the relay channels consist of an amplify-and-forward (AF) relay subchannel and a decode-and-forward (DF) relay subchannel over orthogonal frequency bands. The AF subchannel is used to deliver the computational result shared by user A while the DF subchannel is used to receive the computational tasks required by user $A$. Apparently, user A's offloading strategy could impact the end-to-end delay and the system energy consumption. This work aims to balance these two system performance metric by properly allocating the system computational and communication resources.

Without loss of generality, we assume that the AF relay subchannel and the DF relay subchannel occupy the same width bandwidth $W$. Let $h_{A}^{(1)}$ denote the AF relay subchannel between user A and relay, and $h_{B}^{(1)}$ denote the AF relay subchannel between the relay and user B. Likely, let $h_{A}^{(2)}$ denote the DF relay subchannel between user A and relay, and $h_{B}^{(2)}$ denote the DF relay subchannel between the relay and user B. Moreover, suppose that partial offloading is implemented with \emph{the data partitioned oriented tasks} \cite{110,dvfs}, and the results after computation are proportional to the input size of the tasks. We characterize computation tasks at user A  by the tuple ($L,K,\rho$), where $L$ (in bits) is the size of the tasks before computation, $K$ is the number of required CPU cycles in order to execute each bit, and $0\leq\rho\leq 1$ denotes the data compressed ratio.
 Let $\alpha \in [0,1]$ denote the percentage of computation tasks allocated to the DF relay channel, i.e., $\alpha L$-bits are offloaded to R-MES and $(1-\alpha)L$-bits are computed locally.
In the AF relay subchannel, user A first processes the task locally and transmits the computational results to the R-MES. Hence, the received signal at the relay is
\begin{equation}
y_R^{(1)}=\sqrt{P_{1}^{A}}h_{A}^{(1)} \overline{x_{A1}}+n_R^{(1)},
\end{equation}
where $\overline{x_{A1}}\sim \mathcal{CN}(0,1)$ denotes the the transmit signal after local computing, $n_R^{(1)}\sim \mathcal{CN}(0,\sigma_{R1}^2)$ denotes the complex additive white Gaussian noise (AWGN) at relay, and $P_{1}^{A}$ denotes the transmit power of user A. Next, the R-MES  amplifies the received signal and forwards it to user B. Therefore, the received signal at user B is given by
\begin{align} \label{1}
y_B^{(1)}&{=}\sqrt{P_{1}^{R}}h_{B}^{(1)}y_R^{(1)}+n_B^{(1)}\nonumber\\
&{=}\sqrt{P_{1}^{R}}\sqrt{P_{1}^{A}} h_{B}^{(1)}h_{A}^{(1)}\overline{x_{A1}}{+}\sqrt{P_{1}^{R}}h_{B}^{(1)}n_R^{(1)}+n_B^{(1)},
\end{align}
where $n_B^{(1)}\sim \mathcal{CN}(0,\sigma_{B1}^2)$ and $P_{1}^{R}$ denotes AWGN at user B in the AF relay subchannel and the transmit power of R-MES allocated to the AF relay subchannel, respectively. According to \eqref{1}, the rate and delay in AF relay subchannel can be expressed as
\begin{align}
 \vspace{-0.5em}
R_{\mathrm{AF}}&=\frac{W}{2} \log_2\left(1+\frac { P_{1}^{A}P_{1}^{R}|h_{B}^{(1)} h_{A}^{(1)}|^2}{P_{1}^{R}|h_{B}^{(1)}|^2\sigma_{R1}^2+\sigma_{B1}^2}\right),\\
t_{\mathrm{AF}}&=\frac{(1-\alpha)\rho L}{R_{\mathrm{AF}}}.
 \vspace{-0.5em}
\end{align}
On the other hand, the energy consumption of the AF subchannel is given by
\begin{align}
E_{AF}&=(P_{1}^{A}+P_{1}^{R}|y_R^{(1)}|^2)t_{\mathrm{AF}}\nonumber\\
&=(P_{1}^{A}+P_{1}^{R}P_{1}^{A}|h_{A}^{(1)}|^2+P_{1}^{R}\sigma_{R1}^2)t_{\mathrm{AF}}.
\end{align}


Differently from the AF relay subchannel, in the DF relay subchannel user A first offloads the computational tasks to the R-MES, and then the R-MES decodes the message. Similarly as the AF relay subchannel, we have
\begin{align}
y_R^{(2)}&=\sqrt{P_{2}^{A}}h_{A}^{(2)} x_{A2}+n_R^{(2)},\\
R_{\mathrm{DF1}}&= W\log_2\left(1+\frac{ P_{2}^{A}| h_{A}^{(2)}|^2}{\sigma_{R2}^2}\right),\\
t_{\mathrm{DF1}}&=\frac{\alpha L}{R_{\mathrm{DF1}}},
\end{align}%
where $x_{A2}\sim \mathcal{CN}(0,1)$ denotes the the transmit signal from user A in DF relay subchannel, $n_R^{(2)}\sim \mathcal{CN}(0,\sigma_{R2}^2)$ denotes the AWGN at the R-MES, and $P_{2}^{A}$ denotes the transmit power of user A in the DF relay subchannel.

Then, the R-MES executes edge computing and re-encodes computational results by using the same or a different codebook and forwards the message to user B. After decoding and retransmission user B receives
\begin{equation}\label{3}
y_B^{(2)}=\sqrt{P_{2}^{R}}h_{B}^{(2)} \overline{x_{A2}}+n_B^{(2)},
\end{equation}
where $\overline{x_{A2}}\sim \mathcal{CN}(0,1)$ denotes the the transmit signal after edge computing, $n_B^{(2)}\sim \mathcal{CN}(0,\sigma_{B1}^2)$ denotes the complex AWGN at destination in the DF relay subchannel, and $P_{2}^{R}$ denotes the transmit power of R-MES in the DF relay subchannel.  According to \eqref{3}, the rate and delay from R-MES to destination in DF relay channel can be expressed as
\begin{align}
R_{\mathrm{DF2}}&= W\log_2\left(1+\frac{ P_{2}^{R}| h_{B}^{(2)}|^2}{\sigma_{B2}^2}\right),\\
t_{\mathrm{DF2}}&\overline{}=\frac{\alpha \rho L}{R_{\mathrm{DF2}}}.
\end{align}
Furthermore, the energy consumption of DF relay subchannel communication is given by
\begin{equation}
E_{DF}= P_{2}^{A}t_{\mathrm{DF1}}+P_{2}^{R}t_{\mathrm{DF2}}.
\end{equation}

As in \cite{dvfs}, we model the power consumption of CPU as $P=\eta {F}^3$, where $F$ and $\eta$ are the CPU's computational speed and coefficient depending on chip architecture, respectively. As $F$ is equal to cycles per second, the energy consumption per cycle is thus $\eta {F}^2$. For local computation, its computation energy consumption can be minimized by optimally configuring computational speed via dynamic voltage and frequency scaling (DVFS) technology \cite{dvfs}. When the amount of data bits processed at user A is $(1-\alpha)L$, the execution time $t_l$ is
\begin{equation}
t_{l}=\frac{K_l(1-\alpha)L}{F_{l}},
\end{equation}
where $F_l$ is computational speed of user A. The energy consumption $E_l$ is given by
\begin{equation}
E_l=(1-\alpha) L K_l \eta_l {F_l}^2.
\end{equation}
Similarly, the execution time and the energy consumption of edge computation are given by
\begin{align}
t_{r}&=\frac{K_r \alpha L}{F_{r}},\\
E_{r}&=\alpha L K_r \eta_r {F_r}^2,
\end{align}
where $F_r$ is computational speed of R-MES.

Considering both the AF subchannel and the DF subchannel, the total latency due to executing the whole task is given by
\begin{equation}
t_{\mathrm{sys}}=\mathrm{max} \{ t_{l}+t_{\mathrm{AF}},t_{\mathrm{DF1}}+t_{r}+t_{\mathrm{DF2}}\},
\end{equation}
 and the total system energy consumption is expressed as
\begin{align}
E_{\mathrm{sys}}&=E_l+E_{r}+E_{AF}+E_{DF}.
\end{align}
Consequently, the efficient resource allocation problem can be formulated as

\begin{subequations}
\small{
 \vspace{-1.5em}
\begin{align}
    \mathbf{P1}: \mathop{\min}_{\{\mathbf x\}}&~ E_{\mathrm{sys}}(\mathbf x)+\gamma \cdot t_{sys}(\mathbf x)\label{19a}\\
    \textrm{s.t.} \quad &0 < F_l\leq F_{l,\mathrm{max}},\label{19b}\\
    & 0 < F_r\leq F_{r,\mathrm{max}},\label{19c}\\
    & 0 \leq \alpha \leq 1,\label{19d}\\
       & 0 \leq P_{i}^{A} ,\forall i, \label{19e}\\
      & 0 \leq P_{i}^{R}, \forall i,\label{19f}\\
      & \sum\limits_{i=1} P_{i}^{A} \leq P_{\mathrm{max}}^{A},\label{19g} \\
      & P_{1}^{R}\!\sigma_{R1}^2\!+\!| h_{A}^{(1)}\!|^2P_{1}^{R}P_{1}^{A}+P_{2}^{R} \leq P_{R,\mathrm{max}},\label{19h}
\end{align}
}
\end{subequations}
where $\mathbf x$ denote the set of decision variables $\{\alpha,{P}_{1}^{A},{P}_{2}^{A},$$\\{P}_{1}^{R},{P}_{2}^{R}, F_l,F_r\}$.
The objective function in $\mathbf{P1}$ is a weighted sum of execution delay and system energy consumption with $\gamma $ (in J$\cdot sec^{-1}$) as the weighting factor, which could tradeoff the execution delay and the energy consumption. Constraints \eqref{19b} and \eqref{19c} are the maximum computational speed constraints imposed by user A and R-MES CPU, respectively. Constraints \eqref{19e}, \eqref{19f}, \eqref{19g} and \eqref{19h} specify the transmission power budgets at user A and M-RES.

Problem $\mathbf{P1}$ is nonconvex and nondifferentiable. In addition, the constraint coulping due to \eqref{19h} further complicates problem $\mathbf{P1}$. The popular CCCP algorithm \cite{cccp} can be applied to address problem $\mathbf{P1}$ but incurs a very high computational burden because the CCCP algorithm requires solving a sequence of complex convex optimization problems\cite{CCCP_example}. By exploiting the problem structure, we propose a lightweight iterative algorithm, which is presented in the next section.
 \vspace{-.5em}
\section{Proposed Algorithm}
 \vspace{-.5em}
Observing that the constraints are separable w.r.t the four block variables, i.e., $F_l$, $F_r$, $\alpha$, and $\mathbf{y}\triangleq\{{P}_{1}^{A},{P}_{2}^{A},{P}_{1}^{R},{P}_{2}^{R}\}$, we apply inexact block coordinate descent (BCD) algorithm\cite{IBCD2017} (a variant of BCD algorithm\cite{NonPro1999}) to problem P1. This requires the objective function to be differentiable in general. To address the nondifferentiability issue, we first approximate the objective function of P1 as a smooth function using log-smooth method. Specifically, using the log-sum-exp inequality \cite[pp. 72]{cvx_book} $$\max(x,y){\!\leq\!}\frac{1}{\beta}\log(\exp(\beta x)+\exp(\beta y))\!\leq\!\max(x,y)+\frac{1}{\beta}\log 2$$
we can approximate $t_{\mathrm{sys}}$ as
$$\hat t_{\mathrm{sys}}\approx \frac{1}{\beta}\log(\exp(\beta (t_{l}+t_{\mathrm{AF}}))+\exp(\beta (t_{\mathrm{DF1}}+t_{r}+t_{\mathrm{DF2}})))$$ with a large $\beta$. Hence, the objective of problem $\mathbf{P1}$ can be approximated as a differentiable function, which is given by
\begin{equation}\label{app}
 f_{\beta}(\mathbf{x})= E_{\mathrm{sys}}(\mathbf{x})+\gamma \tilde t_{sys}(\mathbf{x}).
\end{equation}

Now we are ready to use inexact BCD method to solve the smoothed problem, i.e., minimizing $f_{\beta}(\mathbf{x})$ subject to (19b--19h). In the BCD method, each time we update one block variable while fixing the others,  leading to four subproblems.
It can be easily checked that the subproblem w.r.t $F_l$, $F_r$, or $\alpha$ is convex, and thus all these three subproblems can be easily solved using Bisection method\cite{cvx_book}. Therefore, our main efforts are devoted to the update of $\mathbf{y}$.

Let us consider the $\mathbf{y}$-subproblem given by
\begin{equation}\label{p1}
\min_{\mathbf{y}} f_{\beta}(\mathbf{x})~~\textrm{s.t.~~(19e), (19f), (19g), (19h)}.
\end{equation}
Obviously, (19h) is a nonconvex constraint. This makes it difficult to solve the problem. To efficiently update $\mathbf{y}$ while decreasing the objective value,  we apply the concept of CCCP to tackle the nonconvexity of (19h). First, (19h) can be expressed as a DC program:
\begin{align}
\vspace{-0.5em}
\!\!\!\!P_{2}^{R}{+}P_{1}^{R}\!\sigma_{R1}^2\!{+}\!\frac{1}{2}| h_{A}^{(1)}\!\!|^2[(\!P_{1}^{R}\!{+}P_{1}^{A}\!)^2\!{-}\!({P_{1}^{R}}\!)^2\!\!\!{-}\!({P_{1}^{A}}\!)^2]\!{\leq} \! P_{R,\mathrm{max}}\!\label{dc}
\vspace{-0.5em}
\end{align}
By linearizing the nonconvex term $-{P_{1}^{R}}^2-{P_{1}^{A}}^2$ at the current point $\mathbf{\tilde y} =\{\tilde{P}_{1}^{A},\tilde{P}_{2}^{A},\tilde{P}_{1}^{R}, \tilde{P}_{2}^{R}\}$, we approximate \eqref{dc} as a convex constraint
\begin{align}\label{22}
\vspace{-0.5em}
&U(\mathbf{y};\mathbf{\tilde y})\triangleq P_{2}^{R}+P_{1}^{R}\sigma_{R1}^2+\frac{1}{2}| h_{A}^{(1)}|^2[(P_{1}^{R}+P_{1}^{A})^2\!+\!(\tilde{P_{1}^{R}})^2\nonumber\\
&+\!(\tilde{P_{1}^{A}})^2\!-2{\tilde{P_{1}^{R}}}\!{P_{1}^{R}}\!\!-\!2{\tilde{P_{1}^{A}}}{P_{1}^{A}}]\!-\!P_{R,\mathrm{max}}\leq 0 \vspace{-1em}
\vspace{-0.5em}
\end{align}

As a result, we can approximate problem \eqref{p1} as
\begin{equation}\label{p2}
\vspace{-0.5em}
\min_{\mathbf{y}} f_{\beta}(\mathbf{x})~~\textrm{s.t.~~(19e), (19f), (19g), (23)}.
\vspace{-0.5em}
\end{equation}
Now the constraints are all convex. So we can apply \emph{one-step} projected gradient (PG) method\cite{NonPro1999} to problem \eqref{p2},  updating $\mathbf{y}$ according to 
\begin{align}
  \vspace{-0.5em}
    \mathbf{y}^{+}&=P_\Omega[\mathbf{\tilde y}-\nabla f_{\beta}(\mathbf{\tilde y})],\label{23}\\
    \mathbf{y}&=\mathbf{\tilde{y}}+\mu(\mathbf{y}^{+}-\mathbf{\tilde y}),\label{244}
    \vspace{-0.5em}
\end{align}
where $\mu{\in}[0~1]$ can be determined by Armijo rule, $\nabla f_{\beta}(\mathbf{y})$ denotes the gradient of $f_{\beta}$, $\Omega$ denotes the constraint set of problem \eqref{p2}, and $P_\Omega[\cdot]$ denotes the projection of the point $(\mathbf{y}^{+}-\mathbf{\tilde y})$  onto $\Omega$, which is the optimal solution to problem P2
\begin{align}
&\mathbf{P2}: \mathop{\min}_{\{\mathbf y\}}\parallel \mathbf{y}-(\mathbf{\tilde y}-\nabla f_{\beta}(\mathbf{\tilde y})) \parallel^2 \nonumber\\
&\textrm{s.t.} \quad \eqref{19e},\eqref{19f},\eqref{19g},\eqref{22}.
\end{align}

We summarize the proposed BCD algorithm in Table I, where the four block variables are sequentially updated. It can be shown that the algorithm can keep the objective of problem \eqref{p1} nonincreasing and  finally reach a KKT point of problem \eqref{p1}. The proof is omitted due to space limitation.
 \vspace{-1.5em}
\begin{table}[h]
\centering
\caption{Algorithm 1: BCD algorithm for problem \eqref{p1} }
\begin{tabular}{|p{3.1in}|}
\hline
 \vspace{-1em}
\begin{itemize}
\item [0.] initialize the algorithm with a feasible point $\alpha,{P}_{1}^{A},{P}_{2}^{A},{P}_{1}^{R},{P}_{2}^{R}, F_l,F_r$
     \vspace{-1em}
\item [1.]\; \textbf{repeat} \vspace{-1em}
\item [2.]\;\quad update $\alpha,F_l,F_r$ respectively using bisection \vspace{-.5em}
\item [3.]\;\quad  update $\!{P}_{1}^{A},{P}_{2}^{A},{P}_{1}^{R},{P}_{2}^{R}\!$ according to \eqref{23},\eqref{244} \vspace{-1em}
\item [4.]\; \textbf{until} some termination criterion is met\vspace{.5em}
\end{itemize}
\\
\hline
\end{tabular}\vspace{-7pt}
\end{table}

Next we  show how problem $\mathbf{P2}$ can be globally solved using efficient Bisection method. Note that problem $\mathbf{P2}$ is convex. Thus, it can be solved by dealing with its dual problem\cite{cvx_book}. To this end, by introducing Lagrange multiplier $\lambda$ for the constraint \eqref{22}, we define the partial Lagrangian associated with problem $\mathbf{P2}$ as
\begin{align}
\mathfrak{L}(\mathbf{y},\lambda)=\parallel \mathbf{y}-(\mathbf{\tilde y}-\nabla f_{\beta}(\mathbf{\tilde y})) \parallel^2+\lambda U(\mathbf{y; \tilde{\mathbf{y}}}).
\end{align}
Thus, the dual problem of problem $\mathbf{P2}$ can be expressed as
\begin{align}
    \vspace{-1em}
    &\mathop{\max}_{\lambda} h(\lambda)\quad \quad \quad \textrm{s.t.}\quad \lambda\geq0
    \vspace{-2em}
\end{align}
where $h(\lambda)$ is the dual function given by
\begin{equation} \label{dual}
h(\lambda)=\mathop{\min}_{\mathbf y} \mathfrak{L}(\mathbf{y},\lambda)\quad \quad \textrm{s.t.} \quad \eqref{19e},\eqref{19f},\eqref{19g}.\vspace{-1em}
\end{equation}
Note that problem \eqref{dual} can be decomposed into two independent linearly constrained convex quadratic optimization subproblems w.r.t $\{{P}_{1}^{A},{P}_{2}^{A}\}$ and $\{{P}_{1}^{R},{P}_{2}^{R}\}$, respectively, both of which can be globally solved in closed-form. As a result, efficient Bisection method can be applied to problem (31) (and thus $\mathbf{P2}$). The details are omitted due to page limitation.
 \vspace{-.5em}
\section{Numerical Results}
\vspace{-.5em}
In the simulations, all channel gains are modeled as Rayleigh fading with average power loss $10^{-3}$. The variance of complex white Gaussian channel noise in all subchannels is set to $10^{-9}$\cite{2}. The maximum CPU speed of user A and the R-MES is 200 MHz and 600 MHz, respectively. Let $L=1.8\times10^5$ bits, $K_l=K_r=10^3$ cycles/bit, and $\eta_l=\eta_r=10^{-28}$\cite{7}. The maximum transmit power of user A and R-MES is set to 1 Watts and 5 Watts. The smoothness factor $\beta$ is set to $10$.

\begin{figure}[h]
\centering \scalebox{0.35}{\includegraphics {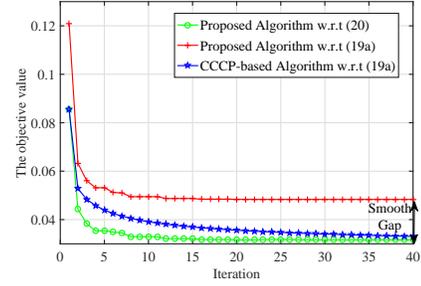}}\vspace{-.5em}
\caption{ Examples of convergence behavior of the proposed algorithm for the case $\gamma=0.01$ and $\rho=0.1$.}\label{fig:f1}
 \vspace{-1.2em}
\end{figure}

First, we show the convergence performance of the proposed algorithm compared with the CCCP method\footnote{The CCCP-based algorithm developed for problem (19) is omitted due to space limitation.} \cite{cccp}. Figure \ref{fig:f1} shows that, despite a gap between the objective values of (19) and (21) due to smooth approximation (see the green line Vs. the red line), the  proposed algorithm can  monotonically converge to the same value as that achieved by the CCCP method (see the green line Vs. yellow line). Furthermore, the proposed algorithm can achieve faster convergence than the CCCP method in terms of the number of iterations. In fact, since the CCCP method requires solving a sequence of complex convex problems, \emph{it takes $60$s on average for convergence in our simulations while the proposed algorithm takes only $0.2$s}. Hence, the proposed algorithm performs much more efficient than the CCCP method.
\vspace{-.5em}
\begin{figure}[h]
\centering \scalebox{0.35}{\includegraphics {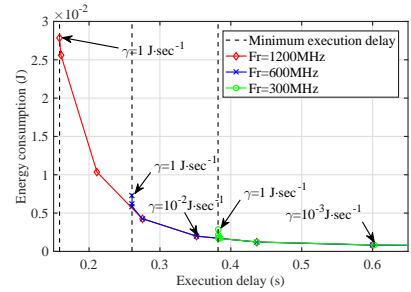}}\vspace{-.5em}
\caption{ Energy consumption vs. execution delay with $\rho=0.1$.\vspace{-.5em}}\label{fig:f2}
\vspace{-.5em}
\end{figure}

Then, the tradeoff between the system energy consumption and the execution delay is shown in Fig. \ref{fig:f2} with $F_{r,\max}=300$, $600$, and $1200$. It can be observed that the energy consumption increases while the execution delay decreases as $\gamma$ increases. Furthermore, when $\gamma$ is relatively large, our design focuses more on delay minimization. As a result, when $\gamma$ increases to some extent, our algorithm could achieve the minimum execution delay (see the dashed line). Conversely, when $\gamma$ is relatively small, our design focuses more on energy consumption minimization and particularly yields the same energy consumption irrespective of the value of $F_{r,\max}$. This is because that the minimum energy consumption is achieved when $F_r$ is very small [cf. (16)].
\vspace{-1em}
\section{Conclusion}
\vspace{-1em}
This paper have considered joint cooperative computation and interactive communication for achieving the minimum weighted sum of the execution delay and the energy consumption in relay MEC systems. For future investigation, it would be interesting to extend this work for two-way relay MEC systems, enabling simultaneous computational results exchange between two users.
\clearpage


\end{document}